\documentclass[titlepage,prd,byrevtex,amsmath,nofootinbib,
preprint,
preprintnumbers,
tightenlines,
]{revtex4-1}

\usepackage{graphicx}
\usepackage{bm}
\usepackage{hyperref}
\newcommand{\beq}{\begin{equation}}
\newcommand{\eeq}{\end{equation}}
\newcommand{\eq}[1]{Eq.~(\ref{#1})}

\begin{document}

\title{Universality of Leading Relativistic Corrections to Bound State Gyromagnetic Ratios}
\author {Michael I. Eides}
\altaffiliation[Also at ]{Petersburg Nuclear Physics Institute,
Gatchina, St.Petersburg 188300, Russia}
\email{eides@pa.uky.edu, eides@thd.pnpi.spb.ru}
\author{Timothy J. S. Martin}
\email{tjmart1@uky.edu}
\affiliation{Department of Physics and Astronomy,
University of Kentucky, Lexington, KY 40506, USA}

%\date{\today}

%\shortauthor{Eides and Martin}

\begin{abstract}

We discuss the leading relativistic (nonrecoil and recoil) corrections to bound state $g$-factors of particles with arbitrary spin. These corrections are universal for any spin and depend only on the free particle gyromagnetic ratios. We explain the physical reasons behind this universality.

%\PACS{12.20.-m, 31.30.J-, 31.30.js}
\end{abstract}

%\keywords{gyromagnetic ratio}

\preprint{UK-10-05}

\maketitle

\section{Introduction}

Gyromagnetic ratios of particles in hydrogenlike bound states have become in the last ten-fifteen years an active field of experimental and theoretical research. The gyromagnetic ratio of a bound electron is proportional to the ratio of the spin flip and cyclotron frequencies of a hydrogenlike ion and to the electron-ion mass ratio. The experimental uncertainties of the ratio of the spin flip and cyclotron frequencies of the hydrogenlike carbon $^{12}C^{5+}$ and oxygen $^{16}O^{7+}$ were reduced to $5-7$ parts in $10^{10}$, see \cite{haffner2003,verdu2004} and review in \cite{mtn2008}. The theoretical expression for the bound state $g$-factor was also greatly improved recently (see, e.g. \cite{pcjy,jent2009} and references in \cite{mtn2008}), and the theoretical uncertainty was reduced to $1.5-5.5$ parts in $10^{11}$. As a result measurements of the bound electron $g$-factor became the best source for precise values of the electron mass in atomic units \cite{mtn2008}. This bright picture is marred by the discrepancy on the magnitude of the leading relativistic corrections to bound state $g$-factors existing in the literature. Leading relativistic (both nonrecoil and recoil) corrections to $g$-factors of loosely bound spin one half particles were calculated a long time ago \cite{gh,fa,co}. These corrections in the case of loosely bound particles with arbitrary spins were calculated in \cite{eg}. It turned out that the leading relativistic and recoil corrections to bound state $g$-factors are universal, they do not depend on the spin of the constituents but only on their free $g$-factors. Among other results, this feature allowed the authors of \cite{eg} to sum all nonrecoil and recoil corrections of order $\alpha^n(Z\alpha)^2$ to bound state $g$-factors in hydrogenlike  ions. Universality of leading binding corrections to bound state $g$-factors was challenged in \cite{mfjetp,fmyaf}. The results of \cite{mfjetp,fmyaf} contained the contributions explicitly depending on spin, both for nonrecoil and recoil corrections. These terms shift the theoretical value of the bound state $g$-factors of the hydrogenlike carbon $^{12}C^{5+}$ and oxygen $^{16}O^{7+}$ by about $2-3$ parts in $10^{11}$. The discrepancy will become even more phenomenologically relevant if proposed improvement \cite{quint2008} of the experimental accuracy by two orders of magnitude is achieved. Later the universal results of \cite{eg} were confirmed in \cite{pach2008,em2010}.

Below we derive an effective nonrelativistic QED (NRQED) Hamiltonian for charged particles with arbitrary spins and calculate the leading relativistic and recoil corrections to the bound state $g$-factors in loosely bound two-particle systems. We show that these corrections are universal for all spins, and explain the physical reasons behind this universality.

\section{NRQED Lagrangian}

A loosely bound two-particle system is effectively nonrelativistic, with characteristic velocities of constituents of order $Z\alpha$. We are looking for the leading nonrecoil and recoil corrections to bound state $g$-factors of order $(Z\alpha)^2$. NRQED is a natural framework for calculation of these corrections. The NRQED Lagrangian sufficient for calculation of these corrections should include terms in nonrelativistic expansion up to and including terms of order $v^2$. The well known (see, e.g., \cite{kn1}) NRQED Lagrangian for the spin one half case is constructed from the covariant derivatives $\partial_0+ieA_0$ and $\bm D=\bm\nabla-ie\bm A=i(\bm p-e\bm A)$, electric and magnetic fields $\bm E$ and $\bm B$, and the spin operator $\bm S$. Notice that in a loosely bound two-particle system the scalar potential$A_0$ is of order $v^2$, $\langle eA_0\rangle\sim(Z\alpha)^2$. For higher spin particles we should, besides the spin, include also higher irreducible intrinsic multipole moments as the building blocks of the NRQED Lagrangian. Technically these  multipole moments are polynomials in the components of the spin operator that for higher spins do not reduce to numerical tensors and operators linear in spin. The general NRQED Lagrangian has the form (compare with \cite{kn1} for spin one half case)

%\begin{FullWidth}
%\LeftColumnBar
\begin{eqnarray} \label{lagr}
{\cal L}=\phi^+\Biggl\{i(\partial_0+ieA_0)+\frac{\bm D^2}{2m}+\frac{\bm D^4}{8m^3}+c_F\frac{e{\bm S}\cdot{\bm B}}{2m}
%\nonumber
%\\
+c_D\frac{e({\bm D}\cdot{\bm E}-{\bm E}\cdot{\bm D})}{8m^2}
\nonumber
\\
+c_Q\frac{eQ_{ij} (D_iE_j-E_iD_j)}{8m^2}
+c_S\frac{ie{\bm S}\cdot({\bm D}\times{\bm E}-{\bm E}\times{\bm D})}{8m^2}
%\nonumber
%\\
+c_{W1}\frac{e[{\bm D}^2({\bm S}\cdot{\bm B})+({\bm S}\cdot{\bm B}){\bm D}^2]}{8m^3}
\nonumber
\\
+c_{W2}\frac{-e D^i({\bm S}\cdot{\bm B})D^i}{4m^3}
%\nonumber
%\\
+c_{p'p}\frac{e[({\bm S}\cdot{\bm D})({\bm B}\cdot{\bm D})
+({\bm D}\cdot{\bm B})({\bm S}\cdot{\bm D})]}{8m^3}\Biggr\}\phi,
\end{eqnarray}
%\RightColumnBar
%\end{FullWidth}

\noindent
where $Q_{ij}=S_iS_j+S_jS_i-(2/3)\bm S^2\delta_{ij}$ is proportional to the electric quadrupole moment operator ($Q_{ij}\equiv0$ for spin one half), and  $\phi$ is a $2s+1$-component spinor field for a particle with spin $s$. We included in the Lagrangian in \eq{lagr} operators of dimensions not higher than four, but omitted some terms (like the terms with derivatives of magnetic field\footnote{See \cite{pach2008} for the explicit form of the Hamiltonian that includes all operators with dimensions not higher than four.}) that are irrelevant for calculation of the leading recoil corrections. Let us mention that gauge invariant bilinears in $\bm E$ and $\bm B$ are of too high order to generate leading relativistic contributions of order $(Z\alpha)^2$ to bound state $g$-factors.

The coefficients in the NRQED Lagrangian for spin one half charged particles are determined from comparison of the scattering amplitudes in nonrelativistic theory and in relativistic QED. We would like to follow the same path for the case of arbitrary spin, but renormalizable QED for charged particles with high spin does not exist. However, the rules for calculation of all one-photon interactions of charged particles with arbitrary spin were constructed some time ago in \cite{kms_pl,kms_jetp}. This construction uses only Lorentz invariance and local current conservation, and is valid for charged particles of arbitrary spin. The interaction vertex that includes all higher multipole moments in the approach of \cite{kms_pl,kms_jetp} is a direct generalization of the ordinary spin one half vertex

\beq \label{vertex}
\Gamma_\mu=e\frac{(p_1+p_2)_\mu}{2m}F_e(q^2,\tau)
-F_m(q^2,\tau)\frac{e\Sigma_{\mu\nu}q^\nu}{2m},
\eeq

\noindent
where $q=p_2-p_1$, $\Sigma_{\mu\nu}$ is the generalization of ordinary spin one half $\sigma_{u\nu}$,  $S_\mu$ is a covariant spin four-vector, $\tau=(q\cdot S)^2$, and $F_e(0,0)=1$, $F_m(0,0)=g/2$. The wave functions are spinors with dotted and undotted indices that are symmetrized among themselves (for more details see \cite{kms_pl,kms_jetp,blp}). The form of the vertex in \eq{vertex} is uniquely fixed by the requirements of Lorentz invariance, $C$, $P$ and charge conservation. Charged particles with higher spins automatically carry higher multipole moments that arise as coefficients in expansion of the form factors $F_e$ and $F_m$ over $\tau$ \cite{kms_pl,kms_jetp}. These intrinsic electric and magnetic multipole moments are treated phenomenologically, and we do not try to calculate them. The phenomenological approach to multipole moments is an advantage for our purposes because we would like to describe how $g$-factors of not necessarily electromagnetic origin (for example the $g$-factor of a spin one deuteron) change in a loosely bound electrodynamic system.

We find coefficients in \eq{lagr} by comparing one-photon scattering amplitudes in NRQED and in the relativistic formalism of \cite{kms_pl,kms_jetp} with vertex \eq{vertex}. Although some terms in \eq{lagr} are bilinear in electromagnetic fields $\bm A$ and $\bm E$ they still  can be restored from one-photon terms due to gauge invariance. Therefore the one-photon relativistic vertex in \eq{vertex} is sufficient for calculation of all the coefficients in \eq{lagr}. We calculated scattering amplitudes off an external electromagnetic field using the nonrelativistic Lagrangian in \eq{lagr} and using the relativistic one-photon vertex in \eq{vertex} at $q^2=0$ and $\tau=0$. In the relativistic calculation we used noncovariantly normalized particle spinors in the generalized standard representation, which is necessary for consistency with the respective nonrelativistic results. Diagrammatically this choice of spinors and representation corresponds to the Foldy-Wouthuysen transformation (for more details, see, e.g., \cite{blp}). After nonrelativistic expansion we compared results of the relativistic calculation with the nonrelativistic ones and obtained values of all constants in the Lagrangian in \eq{lagr}

\begin{eqnarray}
\label{coeff}
c_F=\frac{g}{2},
\quad
c_D=(g-1)\frac{\bm\Sigma^2}{3},
\quad
c_S=g-1,
\quad
c_Q=-2\lambda(g-1),
\nonumber\\
%\quad
c_{W1}=\frac{g+2}{4},
\quad
c_{W2}=\frac{g-2}{4},
\quad
%\nonumber\\
c_{p'p}=\frac{g-2}{2},
%\qquad
\end{eqnarray}

\noindent
where $\bm\Sigma^2=4S$, $\lambda=1/(2S-1)$ for integer spin and $\bm\Sigma^2=4S+1$, $\lambda=1/(2S)$ for half integer spin. Dependence on the magnitude of charged particle spin arose in the coefficients before the Darwin term and the induced electric quadrupole interaction. The $g$-factor in \eq{coeff} is the gyromagnetic ratio defined by the magnetic form factor $F_m(0,0)=g/2$ in \eq{vertex}. In the spin one half case $g$ reduces to a sum of the QED perturbation series if the charged particle is subject only to electromagnetic interactions. The coefficients in \eq{coeff}  in the spin one half case coincide with the respective coefficients in \cite{kn1}, if the phenomenological $g$-factor is substituted in the expressions in \cite{kn1} instead of the perturbative $g=2(1+\alpha/2\pi)$. As an independent test of the effective Lagrangian in \eq{lagr} we considered the charged $W^\pm$-boson sector of the Glashow-Weinberg-Salam Electroweak Theory amended by the anomalous magnetic moment term. We derived the effective NRQED Lagrangian for the $W^\pm$ bosons. This Lagrangian coincides with the Lagrangian in \eq{lagr} for spin one charged particles.

The coefficients in \eq{coeff} are calculated ignoring all loop diagrams and $q^2$ and $\tau$ dependence of the form factors in \eq{vertex}. Both the loop diagrams in relativistic QED and multipole expansion of the form factors would generate further corrections to the coefficients in \eq{coeff}. In ordinary renormalizable spin one half QED (as well as in the renormalizable QED of spin one $W^\pm$ vector bosons) all diagrams, besides those that give contributions only to the free particle $g$-factors, generate corrections to the coefficients that are additionally suppressed by powers of $Z\alpha$. We expect the same effect in any reasonable theory for higher spin particles. It is also obvious that accounting for $q^2$ and $\tau$ dependent terms in the form factors in \eq{vertex} generates terms suppressed by additional powers of $Z\alpha$.  Notice that we ignored the internal electric quadrupole moment implicit in the form factor $F_e(q^2,\tau)$ in \eq{vertex}, but still a term with an induced electric quadrupole moment arose in \eq{lagr} with a coefficient dependent on the magnitude of spin. This immediately means that electric quadrupole interaction with external electric field  depends on the magnitude of spin. Something like this mechanism could in principle make the interaction of the magnetic dipole moment with an external magnetic field spin-dependent, and lead to dependence of bound state $g$-factor on the magnitude of spin. But this happens neither with the coefficient before the term $\bm S\cdot\bm B$, nor with the coefficients before the other terms in \eq{lagr} that give contributions to leading binding correction to the bound state $g$-factors.

Formally two-photon relativistic Compton effect diagrams are needed to obtain coefficients before all terms in the NRQED Lagrangian bilinear in electromagnetic fields. However, as mentioned above, all gauge noninvariant terms in \eq{lagr} bilinear $\bm A$ and $\bm E$  can be restored from one-photon diagrams with the help of gauge invariance. Any gauge invariant terms connected with the two-photon diagrams are of too high order in $Z\alpha$ to contribute to the leading relativistic corrections of order $(Z\alpha)^2$. This allows us to avoid consideration of the relativistic two-photon diagrams.

The Lagrangian in \eq{lagr} with the coefficients from \eq{coeff} is sufficient for calculation of the leading relativistic corrections to the bound $g$-factor in the nonrecoil case.

\section{Effective Two-Particle Hamiltonian. Center of Mass Motion}

Next we construct the effective nonrelativistic quantum mechanical Hamiltonian for a loosely bound electrodynamic system of two particles needed to calculate both nonrecoil and recoil corrections of order $(Z\alpha)^2$ to the bound state $g$-factors.  The interaction between two charged particles with accuracy up to $(Z\alpha)^2$ is described by the one photon exchange which generates Coulomb and Breit interactions. We calculated the two-particle scattering amplitude for particles with arbitrary spins and magnetic moments using the vertices from \eq{vertex} and the photon propagator in the Coulomb gauge. After nonrelativistic expansion we obtained the interaction potential (see also \cite{pach2008,lms_2001})

%\begin{FullWidth}
%\LeftColumnBar
\begin{eqnarray} \label{breit}
V_{int}({\bm p}_1,{\bm p}_2,{\bm r})
=e_1e_2
\biggl[\frac{1}{4\pi r}-(g_1-1) \frac{1}{8m_1^2}\frac{\bm\Sigma_1^2}{3}\delta({\bm r})
-(g_1-1)\frac{3\lambda_1}{\pi}\frac{r^ir^jQ^{(1)}_{ij}}{16m_1^2 r^5}
\nonumber\\
-(g_2-1)\frac{1}{8m_2^2}\frac{\bm\Sigma_2^2}{3}\delta({\bm r})
-(g_2-1)\frac{3\lambda_2}{\pi}\frac{r^ir^jQ^{(2)}_{ij}}{16m_2^2r^5}
-\frac{{\bm r}({\bm r}\cdot{\bm p}_1)\cdot{\bm p}_2}{8\pi m_1m_2r^3}-\frac{{\bm p}_1\cdot{\bm p}_2}{8\pi m_1m_2r}
\nonumber\\
-(g_1-1)\frac{2{\bm S}_1\cdot({\bm r} \times{\bm p}_1)}{16 \pi m_1^2r^3}
+g_1\frac{2{\bm S}_1\cdot({\bm r}\times{\bm p}_2)}{16\pi m_1m_2r^3}
+(g_2-1)\frac{2{\bm S}_2\cdot({\bm r}\times{\bm p}_2)}{16 \pi m_2^2r^3}
\nonumber\\
-g_2\frac{2{\bm S}_2\cdot({\bm r}\times{\bm p}_1)}{16\pi m_1m_2r^3}
+\frac{g_1g_2}{16\pi m_1m_2}\left(\frac{{\bm S}_1\cdot{\bm S}_2}{r^3}
-\frac{3({\bm S}_1\cdot{\bm r})({\bm S}_2\cdot{\bm r})}{r^5}-\frac{8\pi}{3}{\bm S}_1\cdot{\bm S}_2\delta({\bm r})\right)\biggr],
\end{eqnarray}
%\RightColumnBar
%\end{FullWidth}

\noindent
where $\bm r_{1(2)}$, $\bm p_{1(2)}$, $\bm S_{1(2)}$, $m_{1(2)}$, $g_{1(2)}$, and $Q^{(1(2))}_{ij}$ are the coordinate, momentum, spin, mass, gyromagnetic ratio, and induced quadrupole moment of the first (second) particle, and  $\bm r=\bm r_1-\bm r_2$ is the relative coordinate.

This interaction is a natural generalization of the spin one half one-photon potential (see, e.g., \cite{blp}). The only difference is that like in the Lagrangian in \eq{lagr} the coefficients in \eq{breit} before the Darwin terms depend on the magnitude of particles' spins, and new terms with induced electric quadrupole moments arise (we again ignore intrinsic electric quadrupole moments and other intrinsic multipole moments not relevant for calculation of the leading corrections to bound state $g$ factors). The interaction in \eq{breit} is calculated in the absence of a small uniform external magnetic field that is present in the $g$-factor problem. This drawback is easily repaired by the minimal substitution $\bm p_i\to\bm p_i-e_i\bm A_i$, $\bm A_i=\bm B\times \bm r_i/2$.

Combining the nonrecoil Lagrangian in \eq{lagr} and the one-photon potential (after minimal substitution) in \eq{breit} we obtain a total effective nonrelativistic two-particle quantum mechanical Hamiltonian for electromagnetically interacting particles with arbitrary spins (we preserve below only the terms relevant for calculation of the $g$-factor contributions)

\beq \label{totham}
H=H_1+H_2+H_{int},
\eeq

\noindent
where

%\begin{FullWidth}
%\LeftColumnBar
\begin{eqnarray} \label{nonrecham}
H_1=
\frac{({\bm p}_1-e_1{\bm A}_1)^2}{2m_1}
-g_1\frac{e_1}{2m_1}({\bm S}_1\cdot{\bm B})(1-\frac{\bm p_1^2}{2m_1^2})
-(g_1-2)\frac{e_1}{2m_1}({\bm S}_1\cdot{\bm B})\frac{{\bm p}_1^2}{2m_1^2}
\nonumber\\
+(g_1-2)\frac{e_1}{2m_1}\frac{({\bm p}_1\cdot{\bm B})({\bm S}_1\cdot{\bm
p}_1)}{2m_1^2},
\end{eqnarray}
\begin{eqnarray} \label{trexch}
H_{int}=
\frac{e_1e_2}{4\pi r}
+e_1e_2\biggl[-(g_1-1)\frac{2{\bm S}_1\cdot({\bm r} \times({\bm p}_1-e_1\bm
A_1))}{16 \pi m_1^2r^3}
+g_1\frac{2{\bm S}_1\cdot({\bm r}\times({\bm p}_2-e_2\bm A_2))}{16\pi
m_1m_2r^3}
\nonumber\\
+(g_2-1)\frac{2{\bm S}_2\cdot({\bm r}\times({\bm p}_2-e_2\bm A_2))}{16 \pi
m_2^2r^3}
-g_2\frac{2{\bm S}_2\cdot({\bm r}\times({\bm p}_1-e_1\bm A_1))}{16\pi
m_1m_2r^3}\biggr],
\end{eqnarray}
%\RightColumnBar
%\end{FullWidth}

\noindent
and $H_2$ is obtained from $H_1$ by the substitution $1\to2$.

The nonrelativistic effective two-particle Hamiltonian in \eq{totham} describes all (nonrecoil and recoil) leading relativistic corrections to bound state $g$-factors of each of the constituents. The coefficients before all terms in the Hamiltonian in \eq{totham} do not depend on the magnitude of spin, and  the corrections to the bound state $g$-factors are connected with the terms in the Hamiltonian that are linear in external magnetic field. Already at this stage we see that these corrections are universal and do not depend on the magnitude of spin.

Actual calculation of the recoil corrections requires separation of the effects connected with the motion of the bound system as a whole from the internal effects. This task is not quite trivial because the center of mass variables do not separate in the presence of external field. For the current case of a small external magnetic field a solution was suggested in \cite{gh,eg}. The main idea is to impose the condition that the center mass of a loosely bound system moves in a small external field exactly in the same way as a respective elementary particle with the same mass and charge. Neither canonical nor kinetic momentum of a charged particle are conserved in external uniform magnetic field. Instead, in the symmetric gauge $\bm A(\bm r)=\bm B\times\bm r/2$, the pseudomomentum $\bm p+q\bm A(\bm r)$ is conserved \cite{gh},

\beq
[H,\bm p+q\bm A(\bm r)]=0,
\eeq

\noindent
where $\bm p$ and $\bm r$ are the canonical momentum and coordinate of the charged particle, $q$ is its charge\footnote{All charges in this paper carry sign, so, for example, for an electron $e=-|e|$.}, and the Hamiltonian has the standard form $H=(\bm p-q \bm A(\bm r))^2/(2m)$. Classically conservation of pseudomomentum means that the center of the Larmor orbit remains at rest. A naive transition to the standard center of mass coordinates $\bm r=\bm r_1-\bm r_2$, $\bm R=\mu_1\bm r_1+\mu_2\bm r_2$ ($\mu_i=m_i/(m_1+m_2)$) does not secure conservation of total pseudomomentum $\bm P+(e_1+e_2)\bm A(\bm R)$, where  $\bm P=\bm p_1+\bm p_2$. To satisfy the transparent physical requirement of total pseudomomentum conservation (conservation of the position of the Larmor orbit center) transition to the center of mass coordinates should be accompanied by the unitary transformation of the Hamiltonian  $H\to U^{-1}HU$, where $U=e^{i(e_1\mu_2-e_2\mu_1)A(\bm R)\cdot\bm r}$. After this transformation the internal Hamiltonian acquires the form

\beq
H=\frac{\bm p^2}{2m_r}+\frac{e_1e_2}{4\pi r}+H^{(1)}_{spin}+H^{(2)}_{spin}+H_r,
\eeq

\noindent
where $m_r=m_1m_2/(m_1+m_2)$ is the reduced mass, $H^{(1)}_{spin}$ ($H^{(2)}_{spin}$) describes interaction of the first (second) spin with the external field, and $H_r$ includes all other terms in the Hamiltonian.
Explicitly the Hamiltonian for the first particle spin interaction with the external magnetic field  is

%\begin{FullWidth}
%\LeftColumnBar
\begin{eqnarray}
H^{(1)}_{spin} \label{finsping}
=-\frac{e_1}{2m_1}({\bm S}_1\cdot{\bm B})\biggl\{g_1\biggl[(1-\frac{\bm p^2}{2m_1^2})-\frac{e_2[e_1-(e_1+e_2)\mu_1^2]}{24 \pi m_1r}-\frac{e_2[e_2-(e_1+e_2)\mu_2^2]}{12\pi m_2r}\biggr]
\nonumber\\
+(g_1-2)\biggl[\frac{{\bm p}^2}{3m_1^2}
-\frac{e_2[e_1-(e_1+e_2)\mu_1^2]}{24 \pi m_1r}\biggr]\biggr\}.
\end{eqnarray}
%\RightColumnBar
%\end{FullWidth}

\noindent
The Hamiltonian $H^{(2)}_{spin}$ for the second constituent spin has a similar form. The leading binding  correction to the $g$-factor is completely described by the Hamiltonian in \eq{finsping}. We calculate its matrix element with the help of the first order perturbation theory between the Schr\"odinger-Coulomb wave functions that are eigenfunctions  of the unperturbed internal Hamiltonian. After simple calculations we obtain the bound state $g$-factors with account of the leading relativistic corrections of order $(Z\alpha)^2$ for $s$-states with the principal quantum number $n$

%\begin{FullWidth}
%\LeftColumnBar
\begin{eqnarray} \label{g1}
g_1^{bound}=
g_1\biggl[(1-\frac{\mu_2^2e_1^2e_2^2}{2(4\pi)^2 n^2})+\frac{\mu_2e_1e^2_2[e_1-(e_1+e_2)\mu_1^2]}{6 (4\pi)^2 n^2}+\frac{\mu_1e_1e^2_2[e_2-(e_1+e_2)\mu_2^2]}{3(4\pi)^2n^2}\biggr]
\nonumber\\
+(g_1-2)\biggl[\frac{\mu_2^2e_1^2e_2^2}{3(4\pi)^2 n^2}
+\frac{\mu_2e_1e_2^2[e_1-(e_1+e_2)\mu_1^2]}{6 (4\pi)^2 n^2}\biggr],
\end{eqnarray}
\begin{eqnarray} \label{g2}
g_2^{bound}=
g_2\biggl[(1-\frac{\mu_1^2e_1^2e_2^2}{2(4\pi)^2n^2})
+\frac{\mu_1e_1^2e_2[e_2-(e_1+e_2)\mu_2^2]}{6 (4\pi)^2n^2}+\frac{\mu_2e_1^2e_2[e_1-(e_1+e_2)\mu_1^2]}{3(4\pi)^2n^2 }\biggr]
\nonumber\\
+(g_2-2)\biggl[\frac{\mu_1^2e_1^2e_2^2}{3(4\pi)^2n^2}
+\frac{\mu_1e_1^2e_2[e_2-(e_1+e_2)\mu_2^2]}{6 (4\pi)^2n^2 }\biggr].
\end{eqnarray}
%\RightColumnBar
%\end{FullWidth}

These results  resolve the discrepancy mentioned in the Introduction in favor of the results in \cite{eg} (see also \cite{pach2008}). The remarkable property of the expressions in \eq{g1} and \eq{g2} is that they are universal for particles of any spin; they depend only on the $g$-factors of free charged particle, not on the magnitude of their spins.  Technically this happened because no terms in the effective two-particle NRQED Hamiltonian in \eq{totham} relevant for calculation of the leading relativistic corrections contain spin-dependent coefficients $\lambda_i$, $\bm \Sigma^2_i$.

\section{Universality and the Bargmann-Michel-Telegdi Equation}

Universality of the $(Z\alpha)^2$ corrections to the bound state $g$-factors  in \eq{g1} and \eq{g2} requires physical explanation. We already noticed above that simple analysis of the dimensions and spin structure of all terms in the NRQED Lagrangian in \eq{lagr} leads to the conclusion that terms with derivatives of electric fields do not generate contributions to the leading relativistic corrections to the  bound state $g$-factors. Omission of the field derivatives is the basic assumption for validity of the Bargmann-Michel-Telegdi (BMT) equation \cite{bmt,blp}. Hence, one can use the BMT equation for motion of spin in external electromagnetic field for derivation of the leading corrections to the bound state $g$-factors as was suggested in \cite{eg}. Then the leading relativistic corrections are universal if the BMT equation is universal for all spins. To explain universality of the coefficients in the BMT equation let us recall the main steps in its derivation. The basic idea is to use Lorentz invariance to generalize the nonrelativistic precession equation

\beq \label{snorelspinpr}
\dot{\bm S}=g\frac{e}{2mc}{\bm S}\times {\bm B}
\eeq

\noindent
to a relativistically invariant equation for motion of spin in external field. Following  \cite{bmt,blp} we introduce a four-pseudovector $a^\mu$ ($a^2=-\bm S^2$) to describe spin. In the particle rest frame $a^\mu=(0,{\bm S})$. In the relativistic generalization of  \eq{snorelspinpr} the derivative on the LHS should be over proper time $d\tau=\gamma dt$ ($\gamma=1/\sqrt{1-v^2}$). On the RHS we are looking for a vector that is linear and homogenous in the gauge invariant external EM field $F_{\mu\nu}$ and in the spin pseudovector $a_\mu$. Then the most general relativistically invariant equation for spin motion is

\beq
\frac{da^\mu}{d\tau}=\alpha F^\mu\,_\nu a^\nu+\beta u^\mu F^{\nu\lambda}u_\nu a_\lambda,
\eeq

\noindent
where $u^\mu$ is the four-velocity, $u^\mu=(\gamma, \gamma\bm v)$, $u^\mu u_\mu=1$, and $\alpha$ and $\beta$ are unknown constants. The values of these constants are uniquely restored using the nonrelativistic precession equation \eq{snorelspinpr} and the classical relativistic equation of motion for a charged particle in external field. We obtain

\beq
\beta=\frac{e}{m}-\alpha=-(g-2)\frac{e}{2m}.
\eeq

\noindent
Returning to ordinary time $t$ and noncovariant spin vector $\bm S$ we  can write the BMT equation in the form

\begin{eqnarray} \label{noncovbmt}
\frac{d\bm S}{dt}=\frac{e}{2m}{\bm S}\times\biggl\{\left(g-2+\frac{2}{\gamma}\right) \bm B
%\nonumber\\
-\frac{(g-2)\gamma}{1+\gamma}\bm v\cdot\bm B \bm v
+\left(g-\frac{2\gamma}{1+\gamma}\right)[\bm E\times \bm v]\biggr\}.
\end{eqnarray}

\noindent
We see that all coefficients in the BMT equation are universal for any spin, and this universality follows from the universality of nonrelativistic spin precession and Lorentz invariance.

Next we sketch the derivation of the leading relativistic corrections to bound state $g$-factors based on the BMT equation \cite{eg}, in order to demonstrate how the universality of the BMT equation leads to universality of the leading corrections to the  bound state $g$-factors. At the first step we represent \eq{noncovbmt} as the Heisenberg equation of motion for the spin operator

\beq
i\frac{d\bm S}{dt}=[\bm S,H],
\eeq

\noindent
with the Hamiltonian

\begin{eqnarray} \label{bmtham}
H=-\frac{e\hbar}{2m}{\bm S}\cdot\biggl\{\left(g-2+\frac{2}{\gamma}\right) \bm B
\nonumber\\
-\frac{(g-2)\gamma}{1+\gamma}\bm v\cdot\bm B \bm v
+\left(g-\frac{2\gamma}{1+\gamma}\right)[\bm E\times \bm v]\biggr\}.
\end{eqnarray}

To calculate the leading relativistic corrections to the bound state $g$-factor we expand the Hamiltonian in \eq{bmtham} up to quadratic terms in $v^2$, $\gamma\approx 1+\bm v^2/2$.  We obtain the nonrelativistic Hamiltonian

%\begin{FullWidth}
%\LeftColumnBar
\[
H\approx
-\frac{e}{2m}\biggl\{\left(g-\frac{(\bm p-e \bm A)^2}{m^2}\right) {\bm S}\cdot\bm B-(g-2)\frac{((\bm p-e \bm A)\cdot\bm B)({\bm S}\cdot (\bm p-e \bm A))}{2m^2}
\]
\beq \label{bmthamexp}
+\left(g-1\right)\frac{{\bm S}\cdot[\bm E\times (\bm p-e \bm A)]}{m}\biggr\}.
\eeq
%\RightColumnBar
%\end{FullWidth}

\noindent
where we made the substitution $\bm v=(\bm p-e \bm A)/m$ required by gauge invariance. The Hamiltonian in \eq{bmthamexp} coincides with the one in \eq{nonrecham}, and explains why the latter Hamiltonian has universal coefficients.

Besides the Hamiltonian in \eq{nonrecham} calculation of the recoil corrections to bound state $g$-factors also requires knowledge of the interaction Hamiltonian  in \eq{trexch}. Above we derived this interaction Hamiltonian from the relativistic one-photon exchange, but it is easy to see that it includes only the ordinary nonrelativistic spin-orbit and spin-other orbit interactions. It is well known that these interactions are universal for any spin and we could write the Hamiltonian in \eq{trexch} using only nonrelativistic quantum mechanical expressions. This finally explains universality of the corrections of order $(Z\alpha)^2$ to the bound state $g$-factors.

\section{Acknowledgments}
We are deeply grateful to Peter Mohr and Barry Taylor who attracted our attention to the problem of spin dependence. We greatly appreciate the help, advice, and input of Howard Grotch who participated at the initial stage of this project. We thank Krzysztof Pachucki for informing us about his results \cite{pach2008}. This work was supported by the NSF grant PHY--0757928.

%\BalanceColumns[0.5]

\end{document}